\documentclass[longauth]{aa}  

\usepackage{graphicx}
\bibpunct{(}{)}{;}{a}{}{,}
\usepackage{txfonts}
\usepackage{amssymb}
\usepackage{multirow}
\usepackage{float}
\usepackage{gensymb}
\begin{document}

   \title{The SPHERE view of the planet-forming disk around HD100546 \thanks{Based on data collected at the European Southern Observatory, Chile (ESO Programs 095.C-0273(A) and 095.C-0298(A)).}}


   \author{A. Garufi \inst{\ref{ETH}}
   \and S.P. Quanz  \inst{\ref{ETH}}
   \and H.M. Schmid \inst{\ref{ETH}}
   \and G.D. Mulders \inst{\ref{Tucson}, \ref{NASA_Nexus}}
   \and H. Avenhaus \inst{\ref{SANTIAGO}}
   \and A. Boccaletti \inst{\ref{LESIA}}
   \and C. Ginski \inst{\ref{Leiden}}
   \and M. Langlois \inst{\ref{CNRS}}
   \and \\ T. Stolker \inst{\ref{AMSTERDAM}}
   \and J.-C. Augereau \inst{\ref{IPAG}, \ref{Grenoble}}
   \and M. Benisty \inst{\ref{IPAG}, \ref{Grenoble}}
   \and B. Lopez \inst{\ref{Nice}}
   \and C. Dominik \inst{\ref{AMSTERDAM}}
   \and R. Gratton \inst{\ref{Padova}}
   \and T. Henning \inst{\ref{MPIA}}
   \and M. Janson \inst{\ref{MPIA}, \ref{Stockolm}}
   \and F. M\'{e}nard \inst{\ref{UMI}, \ref{IPAG}}
   \and M.R. Meyer \inst{\ref{ETH}}
   \and C. Pinte \inst{\ref{UMI}, \ref{IPAG}}
   \and E. Sissa \inst{\ref{Padova}, \ref{Padova2}}
   \and A. Vigan \inst{\ref{Marseille}, \ref{ESO_Chile}}
   \and A. Zurlo \inst{\ref{SANTIAGO}, \ref{Santiago_Portales}}
   \and A. Bazzon \inst{\ref{ETH}}
   \and E. Buenzli \inst{\ref{ETH}}
   \and \\ M. Bonnefoy \inst{\ref{IPAG}, \ref{Grenoble}}
   \and W. Brandner \inst{\ref{MPIA}}
   \and G. Chauvin \inst{\ref{IPAG}, \ref{Grenoble}}
   \and A. Cheetham \inst{\ref{Geneva}}
   \and M. Cudel \inst{\ref{IPAG}, \ref{Grenoble}}
   \and S. Desidera \inst{\ref{Padova}}
   \and M. Feldt \inst{\ref{MPIA}}
   \and \\ R. Galicher \inst{\ref{LESIA}}
   \and M. Kasper \inst{\ref{IPAG}, \ref{Grenoble}, \ref{ESO_Germany}}
   \and A.-M. Lagrange \inst{\ref{IPAG}, \ref{Grenoble}}
   \and J. Lannier \inst{\ref{IPAG}, \ref{Grenoble}}
   \and A.L. Maire \inst{\ref{Padova}, \ref{MPIA}}
   \and D. Mesa \inst{\ref{Padova}}
   \and D. Mouillet \inst{\ref{IPAG}, \ref{Grenoble}} 
   \and \\ S. Peretti \inst{\ref{Geneva}} 
   \and C. Perrot \inst{\ref{LESIA}}
   \and G. Salter \inst{\ref{Marseille}}
   \and F. Wildi \inst{\ref{Geneva}}
          }

 \institute{Institute for Astronomy, ETH Zurich, Wolfgang-Pauli-Strasse 27, CH-8093 Zurich, Switzerland \label{ETH} \\
              \email{antonio.garufi@phys.ethz.ch}
   \and Lunar and Planetary Laboratory, The University of Arizona, Tucson, AZ 85721, USA \label{Tucson}
   \and Earths in Other Solar Systems Team, NASA Nexus for Exoplanet System Science \label{NASA_Nexus}
   \and Departamento de Astronom\'{i}a, Universidad de Chile, Casilla 36-D, Santiago, Chile \label{SANTIAGO} 
   \and LESIA, Observatoire de Paris-Meudon, CNRS, Universit\'{e} Pierre et Marie Curie, Universit\'{e} Paris Didierot, 5 Place Jules Janssen, F-92195 Meudon, France \label{LESIA} 
   \and Sterrewacht Leiden, P.O. Box 9513, Niels Bohrweg 2, 2300RA Leiden, The Netherlands \label{Leiden}
   \and CNRS/CRAL/Observatoire de Lyon/Universit\'{e} de Lyon 1/Ecole Normale Sup\'{e}rieure de Lyon, Lyon, France \label{CNRS}
   \and Astronomical Institute Anton Pannekoek, University of Amsterdam, PO Box 94249, 1090 GE Amsterdam, The Netherlands \label{AMSTERDAM}  
   \and Univ. Grenoble Alpes, Institut de Plan\'{e}tologie et d'Astrophysique de Grenoble (IPAG, UMR 5274), F-38000 Grenoble, France \label{IPAG}
   \and CNRS, Institut de Plan\'{e}tologie et d'Astrophysique de Grenoble (IPAG, UMR 5274), F-38000 Grenoble, France \label{Grenoble}
   \and Laboratoire Lagrange, Universit\'{e} C\^{o}te d'Azur, Observatoire de la C\^{o}te d'Azur, CNRS, Nice Cedex 4, France \label{Nice}
   \and INAF - Osservatorio Astronomico di Padova, Vicolo dell'Osservatorio 5, 35122 Padova, Italy \label{Padova}      
   \and Max Planck Institute for Astronomy, K\"{o}nigstuhl 17, 69117 Heidelberg, Germany \label{MPIA}
   \and Department of Astronomy, Stockholm University, AlbaNova University Center, 106 91 Stockholm, Sweden \label{Stockolm}
   \and UMI-FCA, CNRS/INSU France, and Departamento de Astronom\'{i}a, Universidad de Chile, Casilla 36-D Santiago, Chile   \label{UMI}
   \and Universit\`{a} degli Studi di Padova, dipartimento di Fisica e Astronomia, vicolo dell'osservatorio 3, 35122 Padova \label{Padova2}           
   \and Aix Marseille Universit\'{e}, CNRS, LAM - Laboratoire d'Astrophysique de Marseille, UMR 7326, 13388, Marseille, France \label{Marseille}
   \and European Southern Observatory, Alonso de Cordova 3107, Casilla 19001 Vitacura, Santiago 19, Chile \label{ESO_Chile}
   \and N\'{u}cleo de Astronom\'{i}a, Facultad de Ingenier\'{i}a, Universidad Diego Portales, Av. Ejercito 441, Santiago, Chile \label{Santiago_Portales}
   \and Geneva Observatory, University of Geneva, Ch. des Maillettes 51, 1290, Versoix, Switzerland \label{Geneva} 
   \and European Southern Observatory, Karl-Schwarzschild-Strasse 2, D-85748 Garching, Germany \label{ESO_Germany}
             }

   \date{Received 10 Dec 2015 / Accepted 19 Jan 2016}

 
  \abstract
   {The mechanisms governing planet formation are not fully understood. A new era of high-resolution imaging of protoplanetary disks has recently started, thanks to new instruments such as SPHERE, GPI and ALMA. The planet formation process can now be directly studied by imaging both planetary companions embedded in disks and their effect on disk morphology. }
   {We image with unprecedented spatial resolution and sensitivity disk features that could be potential signs of planet-disk interaction. Two companion candidates have been claimed in the disk around the young Herbig Ae/Be star HD100546. Thus, this object serves as an excellent target for our investigation of the natal environment of giant planets.}
   {We exploit the power of extreme adaptive optics operating in conjunction with the new high-contrast imager SPHERE to image HD100546 in scattered light. We obtain the first polarized light observations of this source in the visible (with resolution as fine as 2 AU) and new H and K band total intensity images that we analyze with the \textsc{pynpoint} package.}
   {The disk shows a complex azimuthal morphology, where multiple scattering of photons most likely plays an important role. High brightness contrasts and arm-like structures are ubiquitous in the disk. A double-wing structure (partly due to ADI processing) resembles a morphology newly observed in inclined disks. Given the cavity size in the visible (11 AU), the CO emission associated to the planet candidate \textit{c} might arise from within the circumstellar disk. We find an extended emission in the K band at the expected location of  \textit{b}. The surrounding large-scale region is the brightest in scattered light. There is no sign of any disk gap associated to \textit{b}.}
   {}


\authorrunning{Garufi et al.\,2016}

\titlerunning{The SPHERE view on HD100546}

   \maketitle


\section{Introduction}
Our knowledge of the processes governing planet formation will be greatly enhanced by new generation instruments that recently started operations, like the SPHERE \citep[Spectro-Polarimeter High contrast Exoplanet REsearch,][]{Beuzit2008} instrument at the Very Large Telescope (VLT) and GPI \citep[Gemini Planet Imager,][]{Macintosh2014}. The high-contrast and \mbox{-resolution} images enabled by these facilities will continue the recent plethora of observations of planet formation caught in the act. These consist of both the detection of planetary companions still embedded in disks \citep[e.g.,][]{Kraus2012a, Biller2014, Reggiani2014, Sallum2015} and of their imprints on the natal environment \citep[e.g.,][]{Hashimoto2011, Hashimoto2012, Quanz2013b, Garufi2013}. Imaging  disk features (such as disk cavities, spirals, rings etc.) that can be due to the interaction with forming planets is fundamental to determine the framework (and the timeframe) of the planet formation. Excellent first examples of SPHERE's capability to image protoplanetary disks were provided by \citet{Thalmann2015} and \citet{Benisty2015}.   

The disk around HD100546 \citep[B9V star at 97 $\pm$ 4 pc,][]{Levenhagen2006, vanLeeuwen2007} is one of the best laboratories to study the interaction with forming planets since it hosts two embedded companion candidates. The existence of HD100546b (hereafter planet $b$) at $r \simeq 50\  {\rm AU}$ has been proposed by \citet{Quanz2013a}, who detected a bright point source sitting on top of an extended emission in the L$'$ band. This detection was later confirmed by \citet{Currie2014}. \citet{Quanz2015} obtained estimates on its temperature, emitting radius and luminosity, proposing the existence of a warm circumplanetary disk. More recently, \citet{Currie2015} reported the detection of emission in the H band at the location of \textit{b}. On the other hand, the presence of HD100546c (planet $c$) has been claimed by \citet{Brittain2013} via spectroastrometric studies of the CO  and OH ro-vibrational line emission. These authors ascribed the asymmetry of the OH line profiles to gas emission in an eccentric orbit at the disk wall and the annual CO line variability to a concentrated source of emission orbiting the star in proximity of the circumstellar disk wall at $\sim 15$ AU. The latest fit to these variations gives an emitting area of $\sim0.1$ AU$^2$ at $r=12.9\ \rm{AU}$ \citep[when at orbital phase $\phi=6\degree$ from the major axis,][]{Brittain2014}. They noted that this emitting area is only slightly larger than the expected size of a circumplanetary disk around a 5 M$_{\rm J}$ planet at this orbital separation. \citet{Fedele2015} showed that the asymmetric profile of the OH lines can be explained by a misalignment of the spectrograph slit with no need to invoke an eccentric gas disk.

An extensive literature on this remarkable protoplanetary disk exists. The large, roughly $45\degree$-inclined disk was firstly resolved in scattered light by \citet{Pantin2000}. Peculiar disk structures (as the inner cavity, dark lanes and spiral arms) have been resolved in scattered light by \citet{Augereau2001}, \citet{Grady2005}, \citet{Ardila2007}, \citet{Quanz2011}, \citet{Boccaletti2013}, and \citet{Avenhaus2014b}. A quasi-coplanar inner disk at (sub-)AU scale has been studied by \citet{Benisty2010b} and \citet{Tatulli2011} with near-IR interferometry. A rounded disk wall at 11 AU has been claimed by \citet{Panic2014} through mid-IR interferometry. The disk was also resolved with mid-IR imaging \citep{Liu2003} and at millimeter wavelengths by both ALMA and ATCA. The ALMA images (at moderate angular resolution) suggest that the large dust grains are mainly located in form of a ring between the radial locations of planet $c$ and $b$ \citep{Walsh2014, Pineda2014}. \citet{Pinilla2015} modeled these observations and found that if two planets at 10 and 70 AU are responsible for confining the dust in the observed ring, then the outer planet should be at least 2.5 Myr younger than the inner planet. Finally, ATCA observations at 7 mm (with resolution as good as 0.15\arcsec) constrained the disk cavity size to be $\sim$25 AU \citep{Wright2015}. Those observations also reveal a horseshoe-shaped concentration of grains at the disk inner edge.

We present the first SPHERE images of HD100546 in scattered light. The observations reported in this paper consist of high-contrast polarized light images in the visible as well as total intensity images in the near-IR obtained with the SPHERE sub-instruments ZIMPOL and IRDIS respectively. The ZIMPOL observations are the first polarized light images of HD100546 in the visible and are among the highest-resolution {direct imaging data} (${\sim 0.02\arcsec}$) of a protoplanetary disk ever obtained. This paper is organized as follows. In Sect.\ \ref{Observations} we describe observing conditions and data reduction, in Sect.\ \ref{Results} we present the results from both the ZIMPOL and IRDIS images, and in Sect.\ \ref{Discussion} and \ref{Conclusions} we discuss our findings and provide our main conclusions.


\section{Observations and data reduction} \label{Observations}
HD100546 was observed in the context of the Guaranteed Time Observations of the new high-contrast imager SPHERE \citep{Beuzit2008}, operating at the VLT in conjunction with the extreme adaptive optics (AO) system SAXO \citep{Fusco2006}. This paper presents the observations performed with the sub-instruments ZIMPOL \citep[Zurich IMaging POLarimeter,][]{Thalmann2008} and IRDIS \citep[Infra-Red Dual-beam Imager and Spectrograph,][]{Dohlen2008}. Further IFS \citep[Integral Field Spectrograph,][]{Claudi2008} observations were taken along with the IRDIS observations (and will be presented by Sissa et al.\ in prep.). The observing conditions and the reduction of the ZIMPOL and IRDIS data are described in Sect.\ \ref{ZIMPOL} and Sect.\ \ref{IRDIS}, respectively. We also retrieved archival VLT/NACO and Gemini/NICI datasets of HD100546, which are described in Sect.\ \ref{Literature}.

\subsection{SPHERE/ZIMPOL} \label{ZIMPOL}
The ZIMPOL observations were performed on 2015, April 23 (night 1) and May 7 (night 3) in the R$'$ band ($\lambda_{\rm c}= 626$ nm) in Differential Polarization Imaging (DPI). In this mode, ZIMPOL allows polarimetric diffraction-limited observations with very high polarimetric sensitivity. A fast polarization modulator (kHz) provides quasi-simultaneous observations of opposite polarization states, which are registered on the same "even" detector rows during two consecutive modulation cycles. The charges from the first polarization cycle are shifted to the "odd", covered rows and the information about the second state are then collected in the even rows. Furthermore, a half-wave plate (HWP) controls the polarization orientation and permits the observations of a full polarization cycle, consisting of the Stokes parameters $+Q$, $-Q$, $+U$, $-U$ \citep[see e.g.,][]{Tinbergen2005}.  

All our observations were performed in DPI field stabilized mode. We alternated the object orientation by $60\degree$ and dithered its position by 14 pixels on the detector to evaluate any possible instrument artifacts. During night 1, HD100546 was mostly observed in FastPolarimetry mode, which provides the highest polarimetric precision with short exposure time (DIT=1.2 sec). Atmospheric conditions were initially challenging due to the transient presence of thin clouds but converged to good conditions later on (with optical seeing varying from 0.9$\arcsec$ to 2.4$\arcsec$). During the last observing block of night 1 and during all observations of night 3, we operated in SlowPolarimetry (DIT=20 sec) and employed a 155 mas-diameter classical Lyot coronagraph. The weather conditions during night 3 were significantly affected by a strong wind, which resulted in a largely variable seeing (from 0.9$\arcsec$ to 2.2$\arcsec$). A summary of the ZIMPOL observations is given in Table \ref{Settings}.

The DPI data reduction follows the technique illustrated by \citet{Avenhaus2014a}. This includes the equalization of each pair of quasi-simultaneous polarization states to compensate for possible instrumental polarization. This technique assumes the star to be unpolarized. The astrometric calibration was performed following Ginski et al.\ (in prep.), i.e.\ by adopting a plate scale of $3.601 \pm 0.005$ mas/pixel and a detector orientation of $-1.61\degree \pm 0.11\degree$. The final product of our reduction is the pair of polar Stokes parameters ($Q_\phi$, $U_\phi$)\footnote{which was sometimes referred to as ($P_\perp$, $P_\parallel$), ($Q_T$, $U_T$), or ($Q_r$, $U_r$).}. $Q_\phi$ contains the polarized light component tangential to the star on the image plane. $U_\phi$ provides polarization vectors 45$^\circ$ inclined to the tangential component. For face-on systems only, $Q_\phi$ is equivalent to $P=\sqrt{(Q^2+U^2)}$ but it is unbiased (since the signal is not artificially increased by the squares), whereas $U_\phi$ is not expected to contain any signal. In Sect.\,\ref{Discussion_morphology} we discuss the validity of this statement for inclined disks. Since the PSF in many exposures was {degraded} by the high atmospheric turbulence, the final images were produced by stacking a visual selection of best frames (see Table \ref{Settings}). This operation significantly decreased the effective exposure time but also clearly improved the contrast of the final images.

\begin{table}
      \caption[]{Summary of observations. Columns are: night number (see text), observing mode (and waveband), detector integration time (sec) multiplied by number of integrations, number of polarimetric cycles (for ZIMPOL only), and total integration time (sec). Numbers in brackets (for NDIT and P.C.) denote the amount of data used in the final reduction (see text), and the reported $t_{\rm exp}$ reflects this selection.}
         \label{Settings}
     $$ 
         \begin{tabular}{cccccc}
            \hline
            \hline
            \noalign{\smallskip}
            $\#$ & Mode (band) & DIT$\times$NDIT & P.C. & $t_{\rm exp}$ \\
            \hline
            \noalign{\smallskip}
            \multirow{2}{*}{1} & ZIMPOL/FastPol (R$'$) & 1.2$\times$10 & 24 (15) & 720 \\
             & ZIMPOL/SlowPol (R$'$) & 12$\times$6 & 3 (-) & -\\
            \noalign{\smallskip}
            \hline
            \noalign{\smallskip}
            2 & IRDIFS$\_$EXT (K1K2) & 16$\times$256 (235) & & 3760  \\
            \noalign{\smallskip}
            \hline
            \noalign{\smallskip}
            3 & ZIMPOL/SlowPol (R$'$) & 20$\times$6 & 12 (7) & 3360 \\
            \noalign{\smallskip}
            \hline
            \noalign{\smallskip}
            4 & IRDIFS (H2H3) & 16$\times$256 (135) & & 2160 \\
            \noalign{\smallskip}
            \hline
            \hline
         \end{tabular}
     $$ 

   \end{table}

\subsection{SPHERE/IRDIS} \label{IRDIS}
The IRDIS observations were taken in parallel with IFS. This possibility is enabled in IRDIFS mode by dichroic beam splitters and represents the SPHERE nominal infrared mode (with IFS operating at Y-J wavelengths and IRDIS in the H band). Observations in this mode, with IRDIS working in DBI mode \citep[Dual-band imaging mode,][]{Vigan2010} with the H2 and H3 filters ($\lambda_{\rm c}= 1589$ nm and 1667 nm), were taken on 2015, May 28 (night 4). Furthermore, observations in IRDIFS$\_$EXT mode (providing IFS coverage up to the H band and IRDIS working in the K1 and K2, $\lambda_{\rm C} = 2102$ nm and 2255 nm) were taken on 2015, May 3 (night 2). Both runs were carried-out in pupil stabilized mode, allowing us to perform Angular Differential Imaging \citep[ADI,][]{Marois2006}. An apodized pupil Lyot coronagraph with 185 mas-diameter was employed during both runs. During both night 2 and 4, the atmospheric conditions were good, with an average seeing of 0.8\arcsec. The individual frame integration times were 16 seconds and in total 256 frames were obtained in both H and K bands.

The basic data reduction (e.g., bad pixel cleaning, flat fielding, image alignment) was performed using the SPHERE Data Reduction and Handling (DRH) pipeline \citep{Pavlov2008}. From astrometric reference measurements during the same observing run, we derived a pixel scale of $12.210 \pm 0.029$ mas/pixel and a true North orientation of $-1.784\degree \pm 0.129\degree$ \citep{Maire2015}. Visual frame selection was carried-out for both the H and K band data to sort out frames with poor AO correction leading to 121 frames in the H band and 21 frames in the K band that we disregarded from further analysis. The final selection of frames provides a field rotation of $26.8\degree$ in the H band and of $23.3\degree$ in the K band. A summary of the IRDIS observations is given in Table \ref{Settings}.

To subtract the stellar contribution and reveal the circumstellar material, we used the \textsc{pynpoint} package \citep{Amara2012, Amara2015}. \textsc{pynpoint} uses Principal Component Analysis (PCA) to model the stellar contribution in each frame using a basis set of principal components (PCs) that was created from all existing frames. This was done for each filter independently. After stellar subtraction, all frames were rotated to a common sky orientation and mean combined. We varied the number of PCs and the inner/outer radius over which the fit is performed to estimate the impact of these parameters. For the final analysis we eventually settled on simply mean subtracted images (the main PC) as fitting a higher number of PCs led to more flux subtraction and did not reveal any significant additional structure. On the other hand, changing the inner and outer radius of the input images helped to identify features at different spatial scales. In this paper, we present the results obtained with inner/outer diameters equivalent to 0.14\arcsec\ / 1.0\arcsec\ (small scale), 0.5\arcsec\ /  1.5\arcsec\  (medium scale), 1.25\arcsec\ / 5.0\arcsec\  (large scale).  

\subsection{Literature data} \label{Literature}
We also make extensive use of archival data available for HD100546. We retrieved H ($\lambda_{\rm c} = 1630$ nm) and K$_{\rm S}$ ($\lambda_{\rm c} = 2124$ nm) band polarized light images of the source taken with VLT/NACO in April 2006 and in March 2013 \citep[presented by][]{Quanz2011, Avenhaus2014b}. We also recover NACO ADI L$'$ ($\lambda_{\rm c} = 3770$ nm) and M$'$ ($\lambda_{\rm c} = 4755$ nm) band {data} taken in April 2013 and published by \citet{Quanz2015}. Finally, we make use of Gemini/NICI K$_{\rm S}$ band images of HD100546 taken in March 2010 and reduced with the algorithm LOCI \citep{Lafreniere2007} by \citet{Boccaletti2013}.

\section{Results} \label{Results}
The ZIMPOL $Q_\phi$ and $U_\phi$ images resulting from our data reduction are described in Sect.\ \ref{Q_phi} and \ref{U_phi}, respectively. The IRDIS images are illustrated in Sect.\ \ref{Irdis_results}. A comparison between this dataset and the VLT/NACO as well as the Gemini/NICI images is shown in Sect.\ \ref{Comparison_results}.

\subsection{ZIMPOL $Q_\phi$ images} \label{Q_phi}
\begin{figure*}
   \centering
 \includegraphics[width=17.5cm]{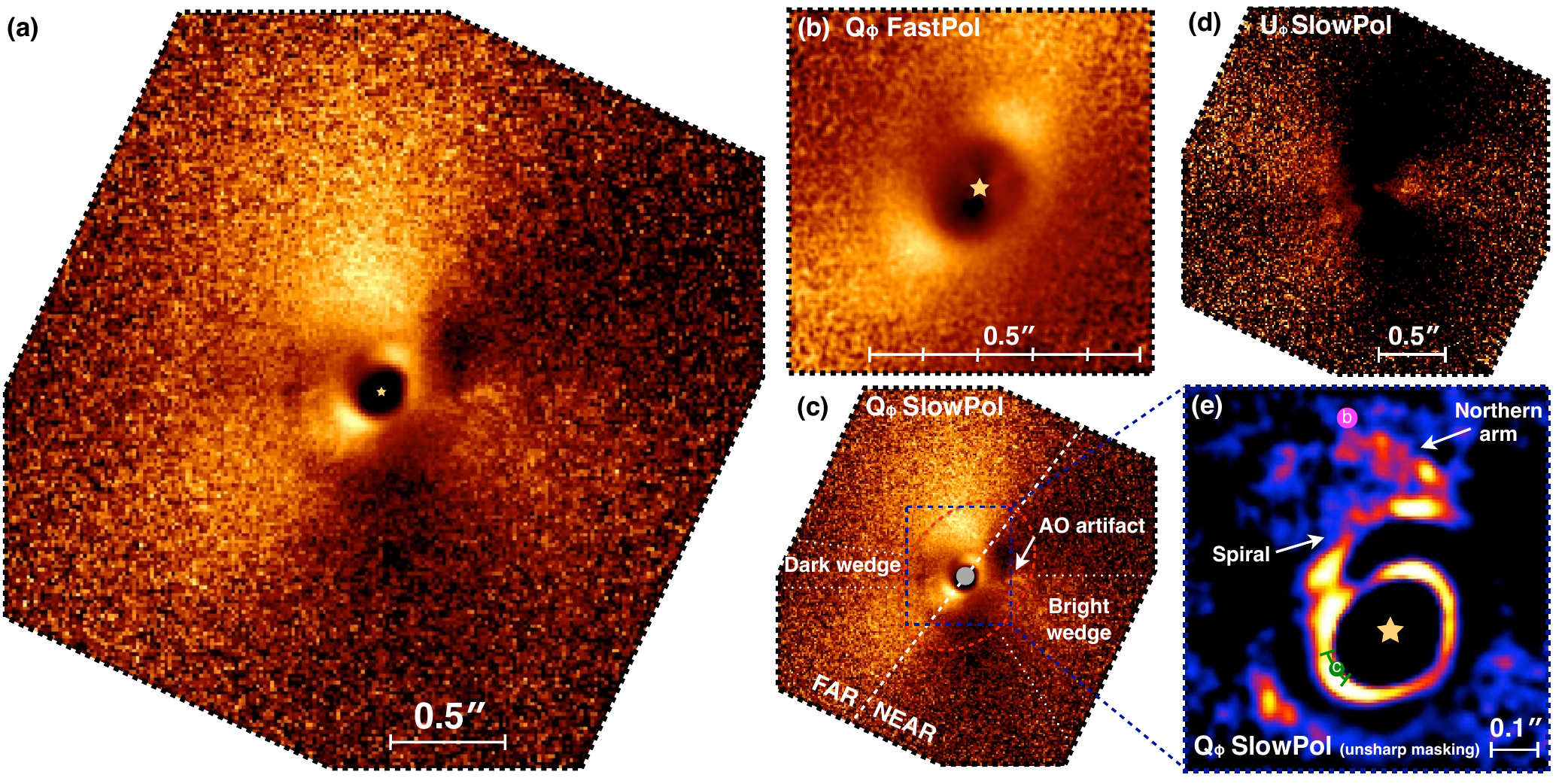}
  \caption{SPHERE/ZIMPOL polarized light imagery of HD100546. \textbf{(a)}: $Q_\phi$ images in coronagraphic SlowPol mode. \textbf{(b)}: $Q_\phi$ images in FastPol mode. \textbf{(c)}: same as (a) with labels, where the white dashed line indicates the disk major axis and the inner grey spot the coronagraph size. \textbf{(d)}: $U_\phi$ images in SlowPol mode, with color stretch twice as hard as in (a). \textbf{(e)}: Unsharp masking of the $Q_\phi$ image (see Sect.\,\ref{structures}). The predicted locations of \textit{b} \citep{Quanz2015} and of \textit{c} in May 2015 with relative azimuthal uncertainty (Brittain S., private comm.) are shown in purple and green. All images except (e) are scaled by the squared distance from the star and are shown with linear stretch. North is up, East is left.}
          \label{Imagery}
  \end{figure*}

\begin{figure*}
   \centering
  \includegraphics[width=9cm]{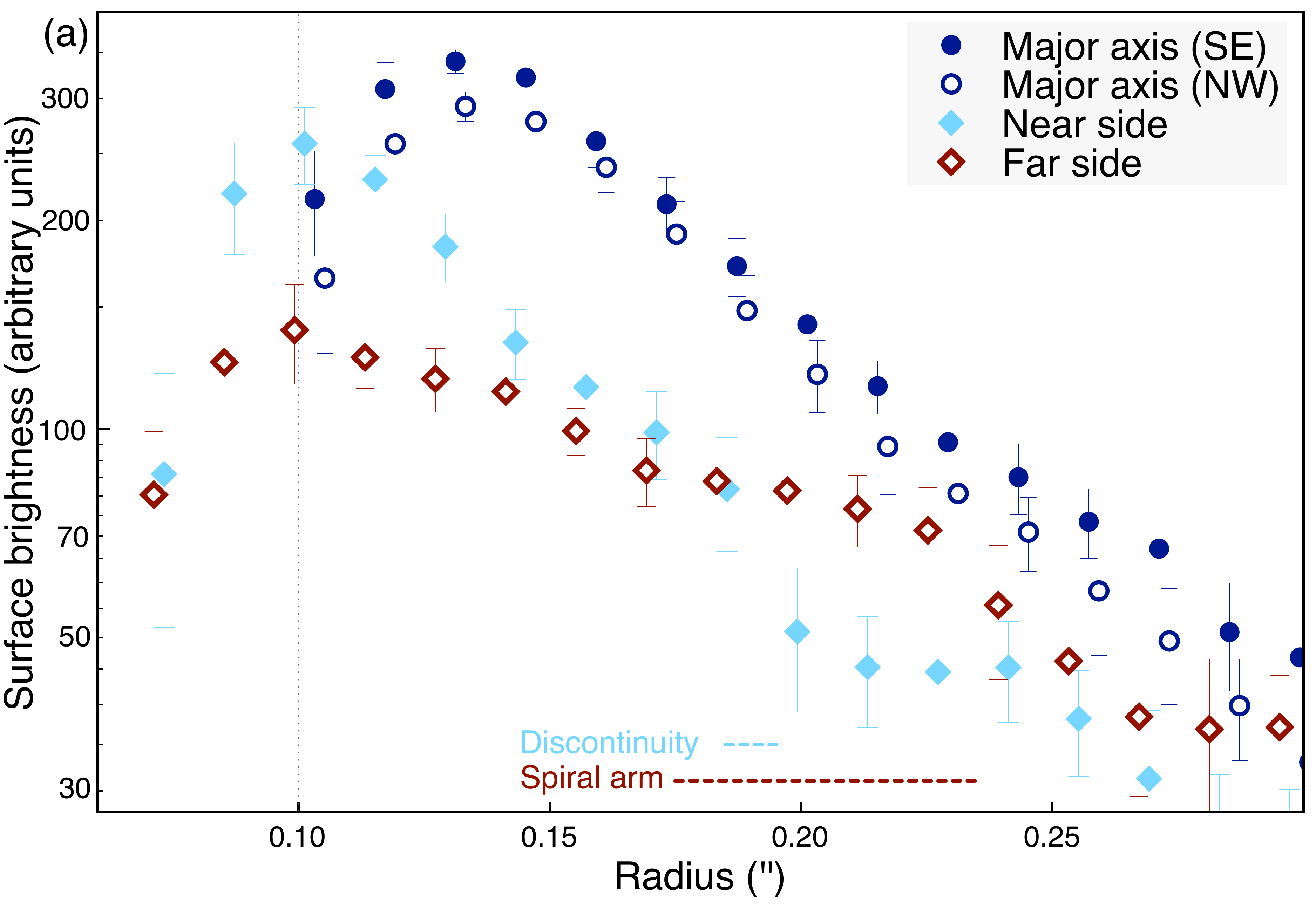}
  \includegraphics[width=9cm]{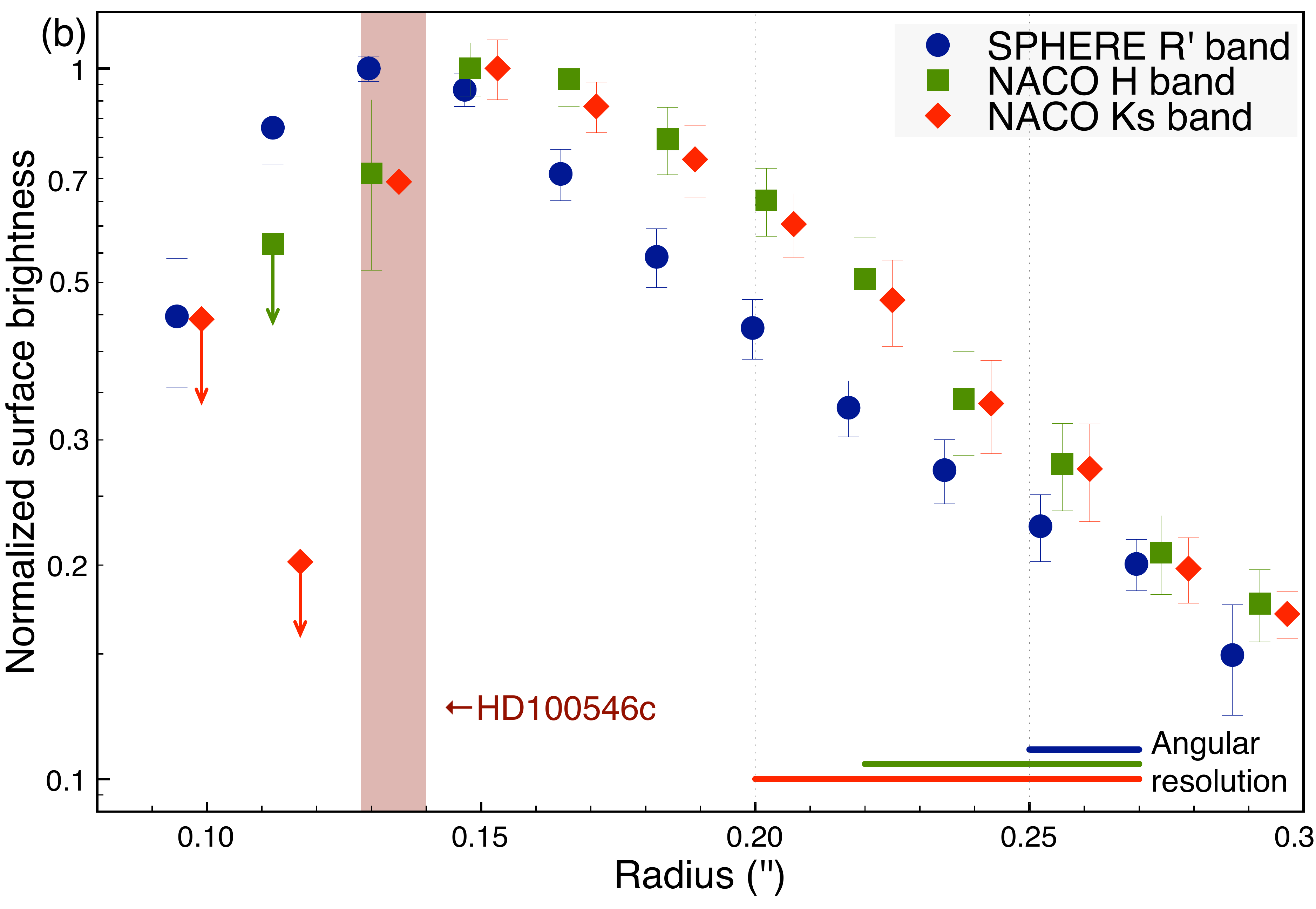}
    \includegraphics[width=9cm]{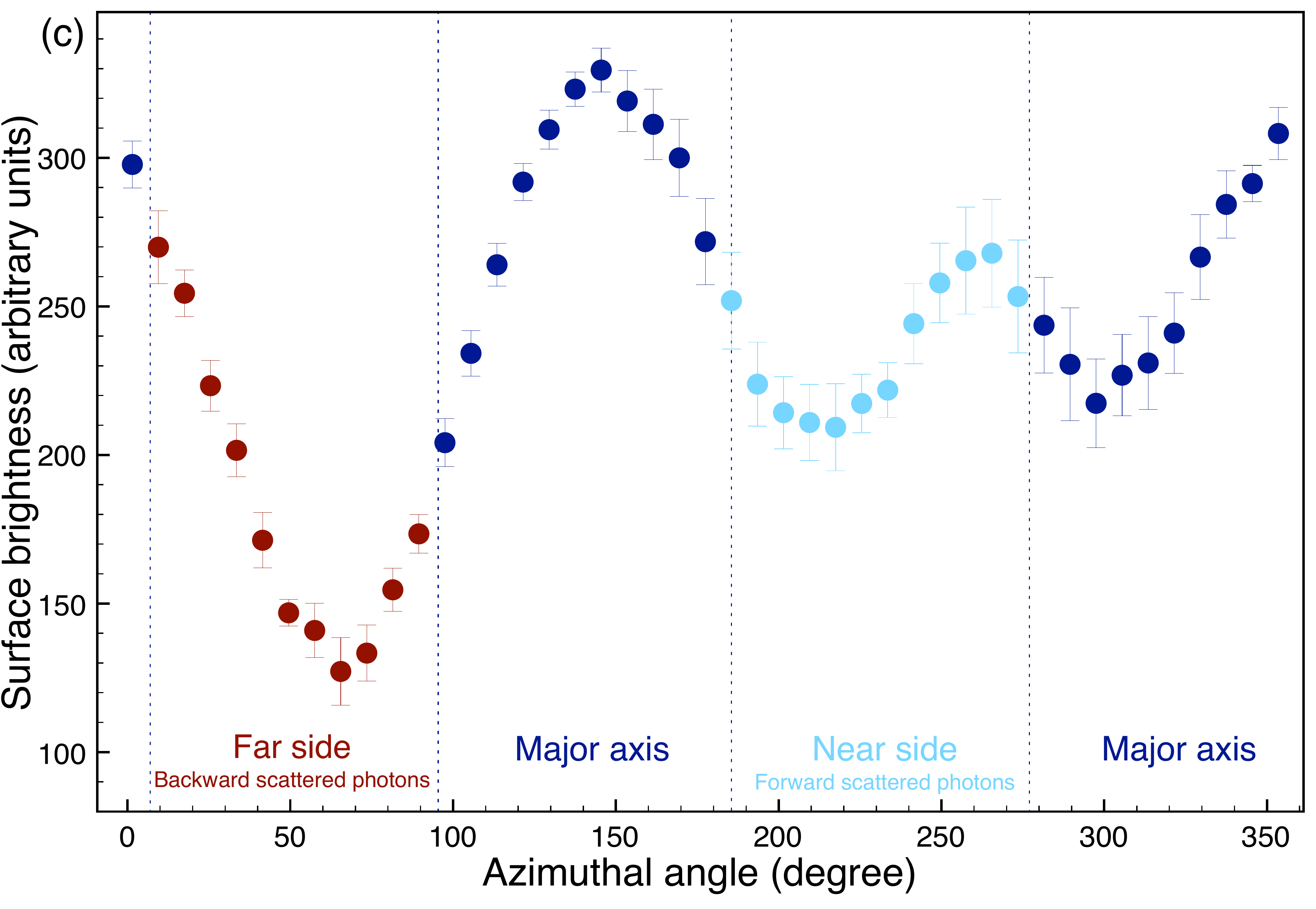}
      \includegraphics[width=9cm]{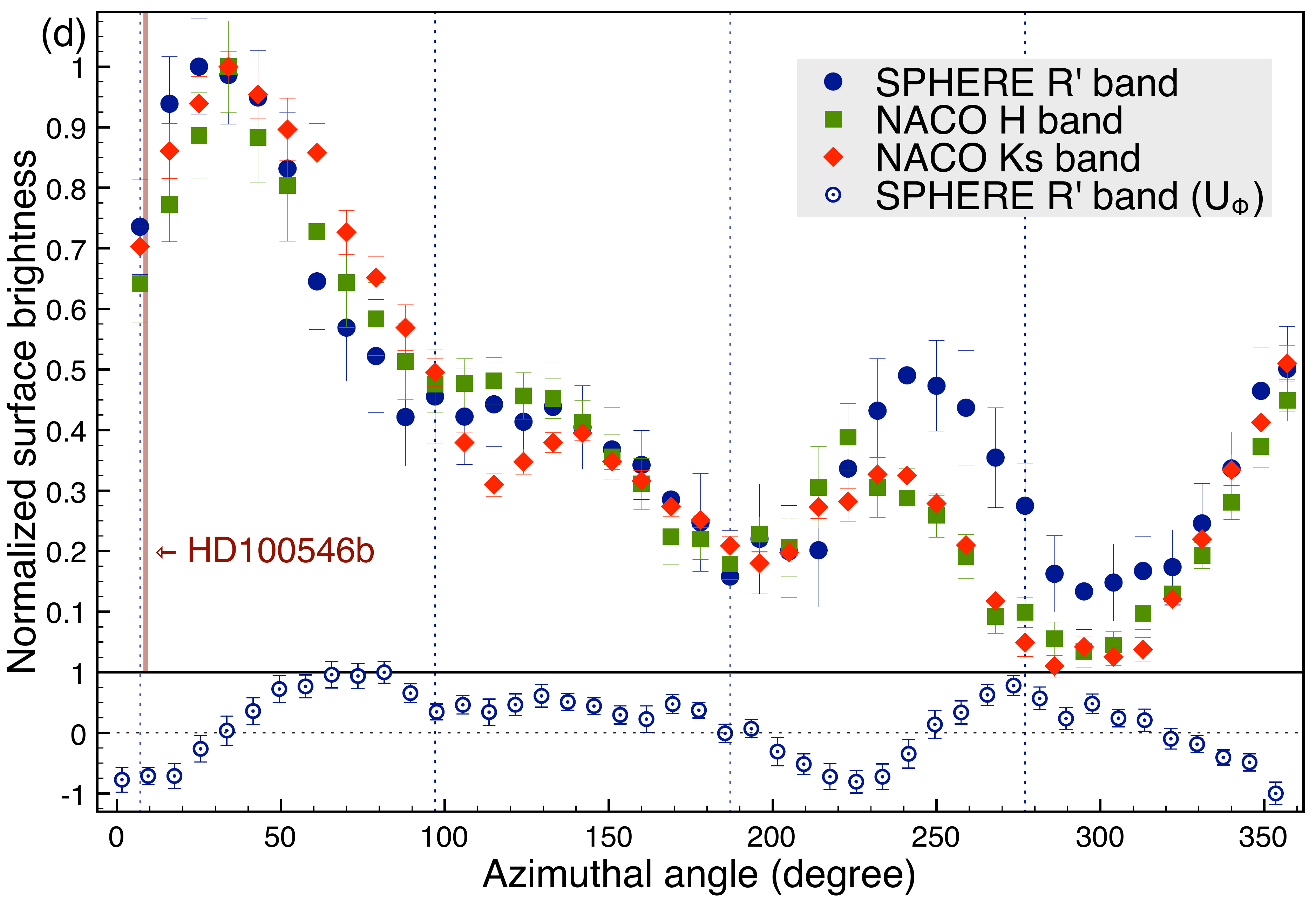}
    \caption{ZIMPOL polarized light brightness profiles. \textbf{(a)}: Radial profile in the inner 0.3\arcsec. The location of the abrupt decrement along the near side axis and the spiral arm bump along the far side axis are highlighted by the horizontal dashed lines. \textbf{(b)}: Radial profile along the SE axis compared to the same from the NACO dataset. Each profile is normalized to the respective peak intensity. The predicted radial location of HD100546c is indicated by the vertical brown area. The angular resolution of the three datasets is shown by the horizontal lines. \textbf{(c)}: Projected azimuthal profile of the inner rim (at $r=0.15\arcsec$ for $i=42\degree$). Angles from North to East. The emission from the four different quadrants is shown with different colors. \textbf{(d)}: Projected azimuthal profile at the predicted radial location of HD100546b (see {red} ellipse in Fig.\,\ref{Imagery}c) compared with the same from the NACO dataset. Each profile is normalized to the respective peak intensity. The azimuthal location of \textit{b} is indicated by the vertical brown region. The vertical dashed lines define the four quadrants as in (c). In the lower space, the same normalized brightness distribution from the SPHERE $U_\phi$ image is shown (note the different y-axis scale therein). All errors are given at 3$\sigma$ level as described in Sect.\,\ref{Q_phi}.}
             \label{Profile}
  \end{figure*}

Figure \ref{Imagery}a and \ref{Imagery}b shows the clear detection of the disk around HD100546 in the $Q_\phi$ images obtained on night 3 and 1 respectively. The images show a globally complex structure. A signal is detected {down to radii as small as} 0.08\arcsec-0.11\arcsec \ (depending on the azimuthal angle), {which is a factor $\sim 4$ larger than the inner working angle of the non-coronagraphic data}. Along the major axis, two bright quasi-symmetric lobes lie outward of the detection inner edge. Farther out, polarized light with an azimuthally complex distribution is detected at radii as large as 0.6\arcsec-1.6\arcsec. Radial and azimuthal brightness profiles from the $Q_\phi$ images are shown in Fig.\,\ref{Profile}. These profiles are obtained by averaging the counts contained in a resolution element (i.e.\ $3\times3$ pixels). Errors are extracted from the $Q_\phi$ images as the standard deviation of the same box divided by the square root of the number of pixels therein. In this section, we analyze the brightness distribution of the $Q_\phi$ images. As from previous interpretations \citep[e.g.][]{Quanz2011}, we will refer to the region close to the star with no detectable signal as the disk cavity (Sect.\ \ref{cavity}) and to the signal at larger radii as the outer disk  (Sect.\ \ref{outer_disk}).  

\subsubsection{Disk cavity} \label{cavity}

The cavity is elongated along the NW-SE direction, with the major axis lying at $\sim140\degree$ East of North. Along this direction, the inner rim reaches its maximum emission on both sides at 0.13\arcsec \ but the brightness of the NW side is $\sim 15\%$ lower (see Fig.\ \ref{Profile}a). Along the SW side \citep[i.e.\ the disk near side,][]{Avenhaus2014b}, the signal is on average $30\%$ lower and the intensity peaks at 0.10\arcsec. This {spatial} difference is consistent with a circular cavity which is $40\degree-50\degree$ inclined. Along the NE side (the far side), the emission is significantly weaker (a median factor 2.3 lower) but the signal also {peaks} at 0.10\arcsec. However, from Fig.\ \ref{Profile}a it is clear that the radial profiles along the near and the far sides are different. Along the near side, the slope is similar to the major axis, except for an abrupt decrease at 0.19\arcsec. On the other hand, the slope along the far side is much more shallow. Furthermore, a brightness \textit{bump} is visible from 0.17\arcsec \ to 0.25\arcsec. This bump can be also appreciated in Fig.\,\ref{Imagery}e as a spiral arm (see Sect.\ \ref{structures}).   

In Fig.\ \ref{Profile}b, we show the ZIMPOL radial profile of the inner 0.3\arcsec \ along the major axis and compare it with the NACO H and K$\rm{_S}$ band profiles. A small radial offset seems to exist between the two datasets. To evaluate this possible discrepancy, we adopted the technique described by \citet{Thalmann2015}. Shortly, we integrated the flux contained in 3-pixel large concentric elliptic annuli, which are obtained from the projection of circular rings \citep[$i=42\degree$ and ${\rm P.A.}=145\degree$,][]{Pineda2014}. We searched for the largest increase in the integrated flux over two contiguous annuli and considered it as the location of the disk inner edge. In parallel, we also generated a disk toy model with surface brightness proportional to $r^{-2}$ outside of a defined radius and only Poisson noise (quantitatively mimicking the image noise) inward. Then, we convolved the synthetic image with a two-dimensional Gaussian with the Full Width Half Maximum (FWHM) of our observations and changed the disk inner edge over a discrete range of values to qualitatively reproduce the signal distribution observed along the major axis.\footnote{The elliptic annuli technique may yield different results by changing the adopted inclination and position angle. However, we did not find any relevant difference over a reasonable range of values. The results from the toy model are valid under the assumption that the disk inner edge is sharp, which is not necessarily true \citep[see][]{Mulders2013b, Panic2014}. However, for the purpose of comparison of the two datasets this uncertainty is not an issue.} Both techniques led to a inner radius of $0.11\arcsec \pm 0.01\arcsec$. For consistency, we applied the same techniques to the NACO datasets and found $0.13\arcsec \pm 0.02\arcsec$ for both H and K$_{\rm S}$ bands. Even though the error bars from these estimates overlap, the small offset from the radial {profiles derived from the two datasets might be real rather than due to the different angular resolution of the observations}.

We also extracted the brightness distribution along the inner rim (Fig.\ \ref{Profile}c) by averaging the contribution from a 5 pixel-wide ellipse at 0.15\arcsec, obtained by projecting a circular ring by ${i=42\degree}$. {We centered the ellipse slightly outside of the intensity peak (0.13\arcsec) to include only regions with detectable signal}. The location of the main peak well matches the major axis. On the contrary, the peak to the NW is $\sim 30\degree$ offset toward North \citep[consistently with the NACO dataset,][]{Avenhaus2014b}. The same amount of offset is found between the intensity peak on the near side and the location of the minor axis.  

\subsubsection{Outer disk} \label{outer_disk}
The azimuthal brightness distribution at large radii is highly asymmetric. The NE half of the image (the far side) is significantly brighter than the SW half (the near side). The distribution on the far side seems to be also discontinuous. As highlighted in Fig.\ \ref{Imagery}c, a dark wedge to East separates two wide bright regions (centered at  $10\degree$ and $120\degree$). On the near side, a bright wedge (spanning $220\degree$-$280\degree$) stands out. A sharp bright blob is also visible to West (marked as AO artifact in Fig.\,\ref{Imagery}c). This is most probably an instrument artifact as it does not appear in differently sky-oriented frames while it is seen in the total intensity images at the radial location of the main AO ring {(corresponding to the AO correction radius at $\approx 20\ \lambda$/D, inside of which the AO system provides almost perfect corrections)}. 

In Fig.\ \ref{Profile}d we show the normalized azimuthal profile from both NACO and SPHERE images, obtained at $r=0.47\arcsec$ \citep[i.e.\ the predicted radial location of \textit{b},][]{Quanz2015} similar to \mbox{Fig.\ \ref{Profile}c}. Interestingly, the intensity peak from the three wavebands lie at the same azimuthal position, which is $\sim 15\degree$ farther East of North of HD100546b. The normalized distributions from the three wavebands are consistent within the error bars almost everywhere. The only exception is the SW bright wedge from the SPHERE image, which is roughly 40\% brighter than the same structures from the NACO datasets. The azimuthal width of this wedge is similar to what we found along the rim (Fig.\ \ref{Profile}c) but it is $\sim20\degree$ displaced.

\subsubsection{Disk structures} \label{structures}
No significant sub-structures can easily be spotted from the outer disk. Only a spiral arm to the NW at $r \sim 0.2\arcsec$ can be seen from Fig.\ \ref{Imagery}a and Fig.\ \ref{Profile}a. To reveal any additional elusive features in the disk, we applied an unsharp masking technique to our images. This technique consists in {adding} a blurred, negative version of the original image to sharpen its details. It has been applied to protoplanetary disks images by e.g., \citet{Ardila2007} and \citet{Quanz2011}. The resulting image must be interpreted with caution as, although sharper, it is {most certainly} not a more accurate representation of the real {disk structure}.

An unsharp masked version of the SPHERE $Q_\phi$ image was obtained by subtracting the smoothed version (by $\sim 10 \times {\rm FWHM}$) from the original image. Since the noise in the science image is high, we firstly subtracted the variance over a FWHM from the original image. {This operation had the effect of decreasing the signal of the inner regions so as to avoid over-subtraction from the smoothed image, and resulted in a much sharper image of the inner 0.5\arcsec.} Finally, the resulting image has been smoothed by one FWHM. The morphology of the final image strongly depends on the choice of the mentioned parameters. However, some features are visible for a broad range of parameters. As shown in Fig.\ \ref{Imagery}e, these are the inner rim, the spiral to NE, and an arm-like structure to North. The reality of this last feature, although not necessarily compelling by itself, will lend credence from the comparison with other datasets (see Sect.\ \ref{Irdis_results} and \ref{Comparison_results}).

\begin{figure*}
   \centering
 \includegraphics[width=18.2cm]{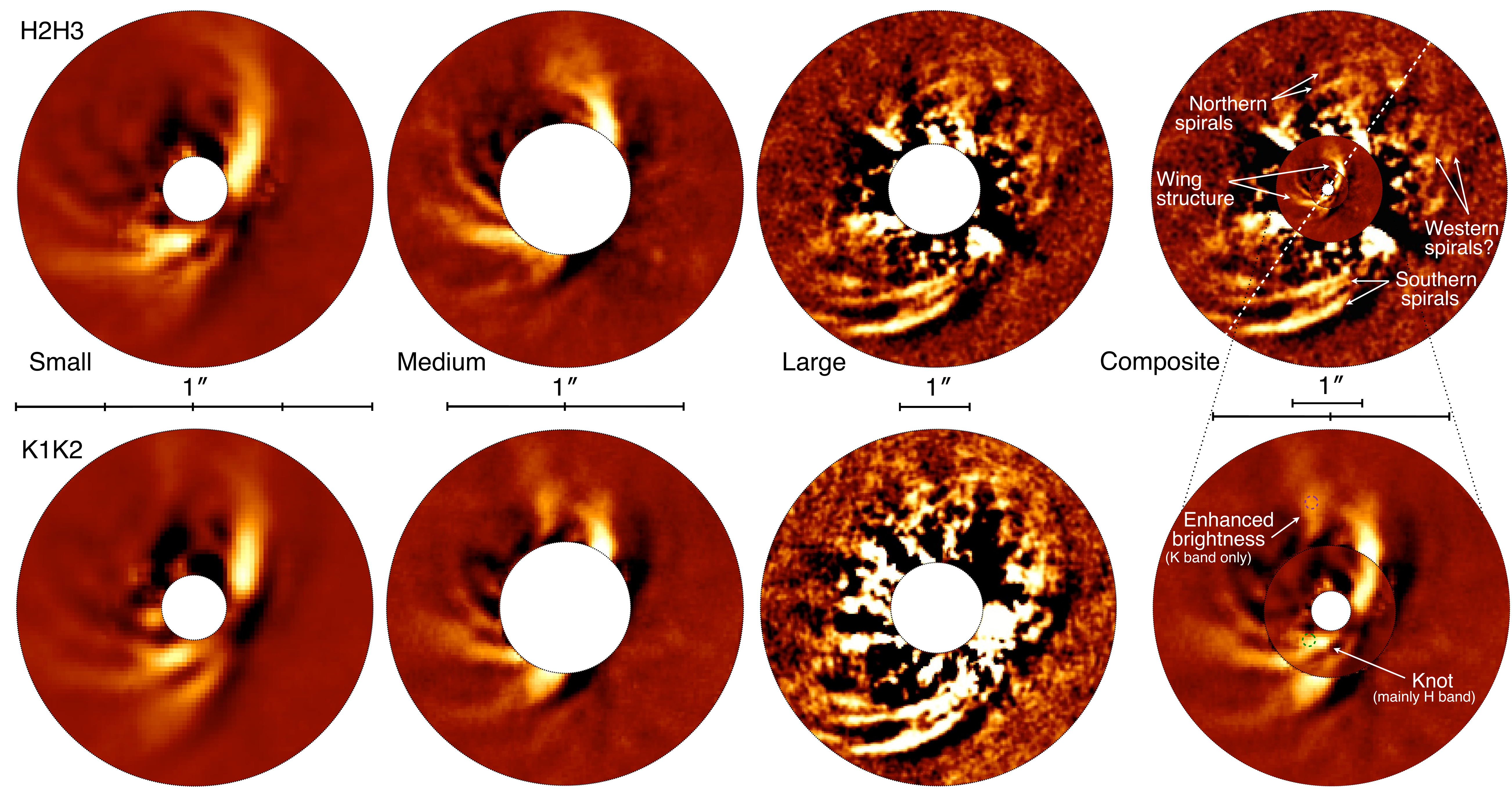}
  \caption{SPHERE/IRDIS imagery of HD100546. Images in the top row are in the H2H3 band, in the bottom row in the K1K2 band. The first three columns are the images resulting from reductions with increasing inner and outer software masks (see Sect.\ \ref{IRDIS}). Images from the third column have been binned by $2\times2$ pixels. The color stretch scales by 20 from first to second column, and by 100 from second to third, while it is arbitrary between the wavebands. The fourth column is a composite image (H band the top one, H/K the small/large bottom one). The dashed line indicates the disk major axis, while the circles point to planet \textit{b} from \citet{Quanz2015} and putative \textit{c} from \citet{Currie2015}. North is up, East is left. }
            \label{Irdis_imagery}
  \end{figure*}

\subsection{ZIMPOL $U_\phi$ images} \label{U_phi}
The $U_\phi$ image resulting from our data reduction is shown in Fig.\ \ref{Imagery}d. This image seems to contain a significant signal pattern. The intensity and the morphology of this signal is not likely to be due to an imperfect data reduction as the equalization of polarization states ({minimizing instrumental polarization,} see Sect.\,\ref{ZIMPOL}) cannot eliminate such a signal and the crosstalk effect between Stokes parameters \citep[see][]{Avenhaus2014a} is a minor effect in SPHERE. Furthermore, the spatial distribution of this signal is consistent through instruments and wavebands \citep[see the $U_\phi$ NACO images by][]{Avenhaus2014b}, with a positive signal being diffusely detected to the East and to the West, and with a negative signal to the NW and SW. Both the bright and the dark wedge highlighted in Fig.\ \ref{Imagery}c seem to have a counterpart in the SPHERE $U_\phi$ image. However, a $\sim 20\degree$ offset exists in the azimuthal location of the bright wedge from the $Q_\phi$ and $U_\phi$ images (it can also be seen in Fig.\ \ref{Profile}d).

More quantitatively, the signal in the SPHERE $U_\phi$ image varies from $\sim 20\%$ of the $Q_\phi$ image at 0.2\arcsec \ to $\sim 35\%$ at 0.8\arcsec. These estimates are obtained by summing up the absolute values of all counts contained in the respective circular annulus. Similar values are found by comparing the peak-to-peak signal at those radii. We will discuss these findings in Sect.\,\ref{Discussion_morphology}. 

\subsection{IRDIS images} \label{Irdis_results}
The IRDIS images resulting from the reduction described in Sect.\ \ref{IRDIS} are shown in Fig.\ \ref{Irdis_imagery}. We found no significant differences between the H2 and the H3 bands as well as between the K1 and the K2 bands. Thus, we only show their combinations (i.e.\ H2H3 and K1K2 - or H and K for simplicity). The overall brightness morphology in the H and K images looks very similar but some minor, though important, differences exist. Different structures are revealed from the reductions with small, medium, and large scales (see Sect.\,\ref{IRDIS}).

\textit{Small scale.} The main features in the small scale images (first column in Fig.\ \ref{Irdis_imagery}) are two bright arms with wide pitch angles (the wings), which are roughly symmetric around the disk minor axis. Signal is detected almost down to the software mask radius at $r\sim 0.10\arcsec$. A bright knot is visible along the SE arm at $r\sim 0.14\arcsec$ (P.A.=$155\degree$) from both wavebands (but it is brighter in H). There is evidence of at least two additional arms to SE with comparable pitch angle. One of these (spanning P.A.\ $90\degree-110\degree$) seems to originate from the bright knot of the SE arm. Interestingly, the spiral arm at 0.2\arcsec \  from the DPI data (see Fig.\ \ref{Imagery}e) is not detected, whereas the faint northern arm matches the location of the Northern wing. 

\textit{Medium scale.} All arms revealed in the small scale images are also visible at medium scale (second column in Fig.\ \ref{Irdis_imagery}). A diffuse enhanced brightness is detected in the K band at P.A.\ $\sim 10\degree$. On top of this extended emission, a more localized emission spans radii from 0.43\arcsec \ to 0.47\arcsec. Interestingly, the region of extended emission roughly lies at the end of the northern arm. We do not detect such an enhanced emission in the H band.

\textit{Large scale.} Many bright arms are also detected in the large scale images (third column in Fig.\ \ref{Irdis_imagery}). The two clearest detections lie to South in both images (the Southern spirals spanning angles $140\degree - 200\degree$). Two additional smaller arms can be seen at P.A. $=140\degree$, close to the easternmost part of the main arms. Moreover, we detect in both bands two similar features to North (the Northern spirals which run in parallel from $40\degree$ to $-20\degree$) and also tentatively two arms which are specular to the Southern spirals around the minor axis (the Western spirals). {Any other structure close to the software mask should not be considered significant, as that region is dominated by subtraction residuals and none of those features is persistent across different frames and reductions.}

\subsection{Comparison with near-IR images} \label{Comparison_results}
\begin{figure*}
   \centering
 \includegraphics[width=18.6cm]{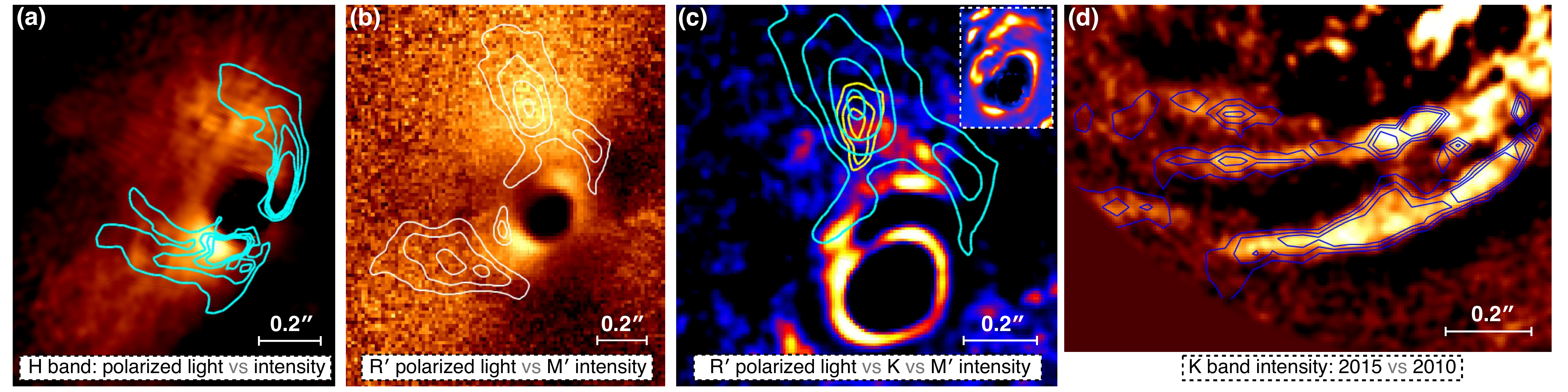}
  \caption{Multi-wavelength and -epoch images of HD100546. \textbf{(a)}: NACO polarized light H band (colors) and IRDIS H band (contours). \textbf{(b)}: ZIMPOL polarized light R$'$ band (colors) and NACO M$'$ band (contours). \textbf{(c)}: ZIMPOL polarized light R$'$ band after unsharp masking (colors), IRDIS K band ({yellow} contours), and NACO M$'$ band ({cyan} contours). The inset image is the NACO polarized light K$_{\rm S}$ band image from 2006 after unsharp masking. \textbf{(d)}: IRDIS K band from May 2015 (colors) and NICI K band from March 2010 (contours).}
            \label{Comparison}
  \end{figure*}

Having collected throughout the years a rich set of data from \mbox{0.6 $\mu$m} to 4.8 $\mu$m, both in polarized light and total intensity, we can obtain insight into both the disk geometry and the properties of scattered light by comparing this image collection. Figure \ref{Comparison} is a selection of these comparisons that we discuss in this section.

Figure \ref{Comparison}a shows the brightness distribution in the H band, highlighting the differences between the polarized light (from NACO, in color) and the total intensity distribution (from IRDIS, contours). It is clear that our IRDIS images can trace the signal as far down as the inner rim in polarized light, but that this is slightly azimuthally offset toward the near side. It is also evident how different the two brightness distributions are at large radii, with the polarized light being more distributed over the angles. Nonetheless, neither of the datasets reveal any significant emission from the SW side, and the locations of the bright/dark boundaries are very well matched. 

In Fig.\ \ref{Comparison}b, the shortest (R$'$) and the longest (M$'$) available wavebands are compared. The two bright large-scale wedges observed in the visible (colors) are also evident from the IR NACO data (contours).  Particularly good is the match between the bright/dark edge to East. Also worth mentioning is the tentative detection of the spiral at 0.2\arcsec \ (Fig.\ \ref{Imagery}e) from the M$'$ dataset.

In Fig.\ \ref{Comparison}c, we show the spatial connection between the northern arm spotted from the ZIMPOL image after unsharp masking (colors) and the diffuse emission from the M$'$ band ({cyan} contours) surrounding planet  \textit{b} \citep{Quanz2015}. The same emission from the IRDIS K band is shown ({yellow} contours). The arm seems to head to the location of \textit{b} and a blurred counterpart of this arm may be seen in the M$'$ band. A minor radial offset between the emission in the K and M$'$ band seems to exist but requires further investigation. The inset image to the top right is the K$_{\rm S}$ band NACO from 2006 \citep{Quanz2011} after applying unsharp masking similarly to the ZIMPOL image. It is clear that the location of the spiral is, to first order, the same.

Fig.\ \ref{Comparison}d is a multi-epoch image of the outer disk region to South. The spiral arms detected by IRDIS in May 2015 are shown in colors while the same features imaged by \citet{Boccaletti2013} in March 2010 are shown in contours. No spatial shift is appreciable from the image.

\section{Discussion} \label{Discussion}
The interpretation of scattered light images is not straightforward. In fact, many physical and geometrical factors may equally contribute to the final appearance of such images. First of all, dust particles scatter photons anisotropically. This scattering angle distribution (referred to as the phase function) depends on the incident radiation as well as on the dust size and shape \citep[see e.g.,][]{Min2012}. {In addition}, scattered photons are polarized also depending on the scattering angle (hereafter the polarizing efficiency). In protoplanetary disks, the scattering angles experienced by the observed photons cannot be precisely determined because of our limited knowledge of the exact disk geometry. Secondly, the amount of radiation incident on the disk may be strongly affected by the geometry of the system at smaller radii (like e.g., warps or belts in proximity of the star). Moreover, multiple scattering on the disk surface may also have an impact on the scattered light distribution. Finally, instrumental contributions or data processing can also significantly alter the final images. 
  
All this might {suggest} caution in interpreting the observed features, as very often these may have twofold (or multiple) explanations. In this section we discuss our findings in the context of different possible scenarios, focussing on the inner disk (Sect.\ \ref{Discussion_cavity}), on the global disk morphology (Sect.\ \ref{Discussion_morphology}), on the disk features (Sect.\ \ref{Discussion_structures}), and on the environment around planet \textit{b} (Sect.\ \ref{Discussion_b}). 

\subsection{Cavity and inner rim} \label{Discussion_cavity}
Our results on the disk cavity (from Sect.\ \ref{cavity}) support the previous findings in DPI \citep{Quanz2011, Avenhaus2014b}, that are that the disk around HD100546 is truncated at roughly 15 AU and that this cavity is consistent with being intrinsically circular. Our data show no sign of offset along the major axis between the center of the cavity and the star, contrarily to what was found by \citet{Grady2005} with spectral observations (${\sim 5~{\rm AU}}$). Along the minor axis, it is harder to claim or rule out any offset because the scattered light distribution from the back side of inclined disks inherently differs from that of the near side. 

\citet{Mulders2013b} showed with mid-IR interferometry that the disk wall is rounded off over a large radial range and that this creates a broad surface brightness profile (10 to 25 AU) peaking at $\sim$12 AU. Our radial profile of Fig.\ \ref{Profile}a may support this scenario. In fact, the brightness distribution along the major axis is constantly decreasing between the peak at 13 AU and 30 AU. This trend differs from those of e.g.\ SAO206462 \citep{Garufi2013} and HD169142 \citep{Quanz2013b}, where an abrupt decline of the brightness profile is seen outward of the intensity peak. However, the exact slope of the brightness distribution at such small radii should not be trusted, because of the PSF smearing effects described by \citet{Avenhaus2014b}.

The radial brightness distribution on the near side is very similar to what is seen along the major axis, except for an abrupt discontinuity (almost 50\%) at a de-projected radius of 25 AU. This may indicate both a shadowed region or a change in the dust distribution at the disk surface. On the other hand, the profile on the far side is completely different. The shallow slope (clear from Fig.\ \ref{Profile}a) may suggest that from 0.1\arcsec \ to 0.2\arcsec \ our observations (partly) trace the disk wall, which is presumably seen at a $\sim 45\degree$ angle from face-on. Such an exquisite data quality may inspire a later, robust modeling of the inner disk geometry to investigate {this} scenario.

Another intriguing result is the possible discrepancy between the location of the disk inner edge in the visible (11 AU) and in the near-IR (13 AU). This incongruity, though marginal, may be the result of a gradual increase of optical depth at the inner rim with disk opacity $\kappa _\lambda$ varying with the wavelength. To {explore} this scenario, we calculated the spectral index $\beta$ of ${\kappa _\lambda \propto \lambda ^\beta}$ necessary to explain this discrepancy. We imposed the observed 11 AU and 13 AU {edges} as the $z_1$ yielding $\tau(z) \equiv \int_{z_0}^{z_1} \kappa_\lambda dz=1$ in the visible and near-IR respectively, for a range of realistic values of $z_0$. This exercise resulted in $\beta$ values spanning from -0.5 to -1.3, which is consistent with the expectations from \mbox{(sub-)$\mu$m} sized particles \citep[e.g.,][]{Backman1993}. We note that the disk inner edge inferred from the visible dataset matches the location of the bright ring of emission ($11 \pm 1$ AU) claimed through mid-IR interferometry by \citet{Panic2014}.  

Interferometric ATCA images of HD100546 \citep{Wright2015} suggest that the cavity size at millimeter wavelengths is roughly 25 AU, which is more than twice as large as our estimate in the visible. Such a large discrepancy is a probable indication of the different behavior of $\mu$m- and mm-sized dust grains in gapped disks. In fact, many authors have shown that a pressure bump at the disk inner edge may act as a filter, allowing grains smaller than a certain size to drift inward but efficiently trapping larger grains \citep[e.g.,][]{Rice2006, Brauer2008, Dong2012}. \citet{Pinilla2012} have shown that two distinct cavity sizes are expected from the interaction of a disk with a planet, one for the gas (closer to the companion) and one for the millimeter particles farther out. This morphology has been observed in a number of disks by comparing scattered light images (tracing $\mu$m-sized grains which are well coupled to the gas) and millimeter interferometric images \citep[e.g.,][]{Garufi2013, Tsukagoshi2014, Follette2015}. More specifically, \citet{deJuanOvelar2013} simulated visible and millimeter images of disks with a cavity carved by planets with different masses and orbital radii. Their results provide a fitting function to relate at different radii the visible/millimeter cavity size ratio with the planet mass. In the case of HD100546, this ratio is observed to be $\sim$0.5. Extending their results to $r < 10$ AU, one finds that a companion responsible for this dust differentiation should be more massive than 15 ${\rm M_{jup}}$. Adopting {an} analytic solution to this difference as in \citet{Garufi2013} \citep[after the model by][]{Pinilla2012} leads to an even higher lower limit ($\sim 40 \ {\rm M_{jup}}$). Similar values were estimated by \citet{Mulders2013b} from the aforementioned shape of the disk wall and by \citet{Pinilla2015} by modeling ALMA observations. The scenario where multiple planets contribute to sculpt the disk and to differentiate the dust grains, though a valid explanation, has not been extensively studied because of the degeneracies {introduced}.

Given these considerations, the potential companion \textit{c} associated with the compact CO emission by \citet{Brittain2013} is unlikely to be the (unique) cause of the observed cavity. From Fig.\ \ref{Profile}b, it is clear that such an object would lie within the \mbox{$\mu$m-sized} particles halo, disfavoring a scenario where these particles are retained at 11 AU because of the interaction with this object. Furthermore, the size of this CO emission would be associated with a planet with a relatively low mass \citep[$\sim 5 \ {\rm M_{jup}}$,][]{Brittain2014}, which is not consistent with the large cavity size nor with the observed $\mu$m-mm dust differentiation in the framework of the current models. An intriguing possibility is that this object is actually a consequence of the observed cavity, since the accumulation of material is favored by the pressure bump at the disk inner edge \citep[e.g.,][]{Pinilla2012}.

One would be tempted to relate the location of \textit{c} in May 2015 (see Fig.\ \ref{Imagery}e) with both the increased brightness of the SE side of the rim and the possible starting point of the main spiral. However, these {associations} may be fortuitous as both features were morphologically similar in the NACO dataset from April 2006 \citep[see also Sect.\ \ref{Discussion_structures}]{Avenhaus2014b}, when the companion candidate was orbiting at angles close to North \citep{Brittain2013}. Recent H band images of HD100546 suggest the putative detection of \textit{c} at angles of $150\degree$ \citep{Currie2015}. The location of this detection roughly matches the bright knot of the SE arm from the IRDIS images at the convergence of another arm (see bottom right panel of Fig.\,\ref{Irdis_imagery}). However, the morphological similarity with the NW arm (both in brightness and symmetry around the minor axis) and the absence of such a knot in the K band images may suggest that the signal detected at those locations is pure disk emission. This discourages us from further analyzing and interpreting this feature. All in all, our dataset does not firmly reveal the presence of planet \textit{c} nor of any disk feature which may be connected to the interaction with the planet candidate.  

\subsection{Disk morphology and scattering properties} \label{Discussion_morphology}
The complex azimuthal distribution of the scattered light from HD100546, from both ZIMPOL and IRDIS, can only partially be explained by the scattering phase function. In this section, we discuss what other factors may contribute to the observations.

\textit{$Q_\phi$ vs $U_\phi$ images.} The presence of some diffuse features of the $Q_\phi$ image (like the bright and dark wedges) also in the $U_\phi$ images (see Fig.\ \ref{Imagery}d) may cast doubt on their intrinsic existence. In fact, the $U_\phi$ parameter has largely been used to determine the noise level of polarimetric observations \citep[e.g.,][]{Garufi2013, Avenhaus2014b}, since by construction it was not thought to contain any scattering information. However, the amount ($20\%$ to $35\%$ of the $Q_\phi$ images) and the consistency (between different wavebands and instruments) of the $U_\phi$ image of HD100546 brought this view into question. \citet{Bastien1988} showed that in optically thick, inclined disks the effect of multiple scattering may lead to significant deviations from purely azimuthal linear polarization. Recently, \citet{Canovas2015} have produced synthetic $U_\phi$ images of disks with different inclinations and masses, showing that these images contain a strong signal in disks with $i>40\degree$. The morphology of the synthetic images resemble that of HD100546, with the near side of the disk showing an alternation of strongly negative and positive signal. All this may suggest that multiple scattering in the disk of HD100546 may act to \textit{transfer} a fraction of the polarized signal from the $Q_\phi$ to the $U_\phi$ image and therefore that the presence of features in both images does not necessarily discredit their intrinsic existence.

\textit{Inner rim.} From Fig.\ \ref{Imagery}a and Fig.\ \ref{Profile}c, it is clear that the polarized emission at the inner rim is maximized along the major axis. An analogous morphology is observed in other inclined disks \citep[e.g.,][]{Hashimoto2012, Garufi2014b}. This is not surprising because the polarizing efficiency of scatterers is typically maximized for scattering angles around $90\degree$ \citep[e.g.,][]{Murakawa2010}. It is also clear from Fig.\ \ref{Comparison}a that the (non-polarized) emission from IRDIS is maximized at angles somewhat closer to the minor axis. This indicates that dust grains at the inner rim preferentially scatter in a forward direction. However, the net amount of this trend must be small compared to the polarizing efficiency, so as to maximize the polarized light (which is a combination of phase function and polarizing efficiency) along the major axis. 

\textit{Polarized light from the (bright) far side.} At larger radii ($> 0.3\arcsec$), the polarized emission is predominantly distributed on the disk far side. This morphology can be explained by the disk geometry. In fact, {for} inclined and flared disks (like HD100546) the scattering angles experienced by photons from the back side are closer to $90\degree$. However, the high brightness contrast between the near and the far side and the darkness of the near side also in the IRDIS (non-polarized) images may both suggest backward-scattering particles at the disk surface \citep[see also work by][]{Avenhaus2014b}. Specifically, two brighter regions are clear (at angles around $10\degree$ and $120\degree$) and these are also visible in the M$'$ band (see Fig.\ \ref{Comparison}b). These regions are offset from the major axis by a different angles respectively ($52\degree$ to North and $28\degree$ to SE). This may indicate a larger disk flaring angle to North, since higher disk scale heights \textit{move} the $90\degree$-scatters toward the far minor axis\footnote{One can relate the scattering angle $\theta$ to the disk geometry through $\theta=90\degree+\sin{\rm (P.A.)}\cdot i - \beta$ with $i$ disk inclination, P.A.\ position angle from the major axis, and $\beta$ disk opening angle.}. This scenario is consistent with the different brightness of the two regions (with the North being twice as bright) and would also explain why the offset with the major axis increases with the radius ($\sim 10\degree$ at the rim and up to $50\degree$ for higher disk scale heights). On the other hand, the dark wedge to East (see Fig.\ \ref{Imagery}c) is too localized to be described by a change in the disk scale height. A possible explanation for this deficit is a shadow. Shadows in protoplanetary disks have been predicted \citep[e.g.][]{Dullemond2001} and observed \citep[e.g.][]{Avenhaus2014a}. The compact inner dust belt \citep[e.g.,][]{Benisty2010b, Mulders2013b} may be responsible for such a shadow, similar to HD142527 \citep{Marino2015}. 
An alternative cause could be the spiral at 20 AU (see Fig.\ \ref{Imagery}e), whose inner (brighter) portion is azimuthally consistent with the dark wedge. In this scenario, the dark wedge should rotate with the spiral (see Sect.\ \ref{Discussion_structures}).   

\textit{Polarized light from the (dark) near side.} The dark region to {the} SW is also challenging to interpret. This discussion must be related to the interpretation of the bright wedge therein (see Fig.\ \ref{Imagery}c) which is $40\%$ brighter than in the NACO dataset (Fig.\ \ref{Profile}d). \citet{Avenhaus2014b} suggested that the dark region is the result of a particularly backward-peaked scattering phase function (where photons from this region are scattered by angles as small as $50\degree-70\degree$). In this scenario, the bright wedge could represent the low tail of a forward-scattering peak, which is typically estimated to be {in the range} $0\degree-30\degree$ \citep[e.g.,][]{Min2016}. These scattering angles are typically not observed in moderately inclined disks. However, a combination of high flaring angle and broad peak of the phase function may enable this in HD100546. Alternatively, this darkness may also be due to a large-scale shadow. \citet{Wright2015} revealed from millimeter observations a horseshoe structure lying at the SW disk inner edge and claimed that such an accumulation of particles may shadow the outer disk. However, our data trace the dark region down {to} 20 AU whereas the inner edge of the millimeter emission by \citet{Wright2015} is {measured to be} 25 AU. Furthermore, the presence of the bright wedge (highlighted in Fig.\ \ref{Imagery}c) is difficult to reconcile with this scenario. It could be a region of penumbra due to asymmetries in the millimeter horseshoe, but the presence of this bright wedge stretching down to the disk inner rim (see Fig.\ \ref{Profile}c) disfavors this view. Nonetheless, these arguments do not rule-out a shadow cast by the aforementioned inner dust belt. A good vehicle to solve this ambiguity is inferring whether the different intensity between the SPHERE (visible) and the NACO (near-IR) datasets is due to the different epochs (which would favor the shadow scenario) or to the different wavebands (which would suggest a small dependency {of} the phase function on wavelength). Therefore, new near-IR observations (with e.g., SPHERE/IRDIS in DPI mode) will probably clarify the nature of the dark region.

\begin{figure}
   \centering
 \includegraphics[width=9cm]{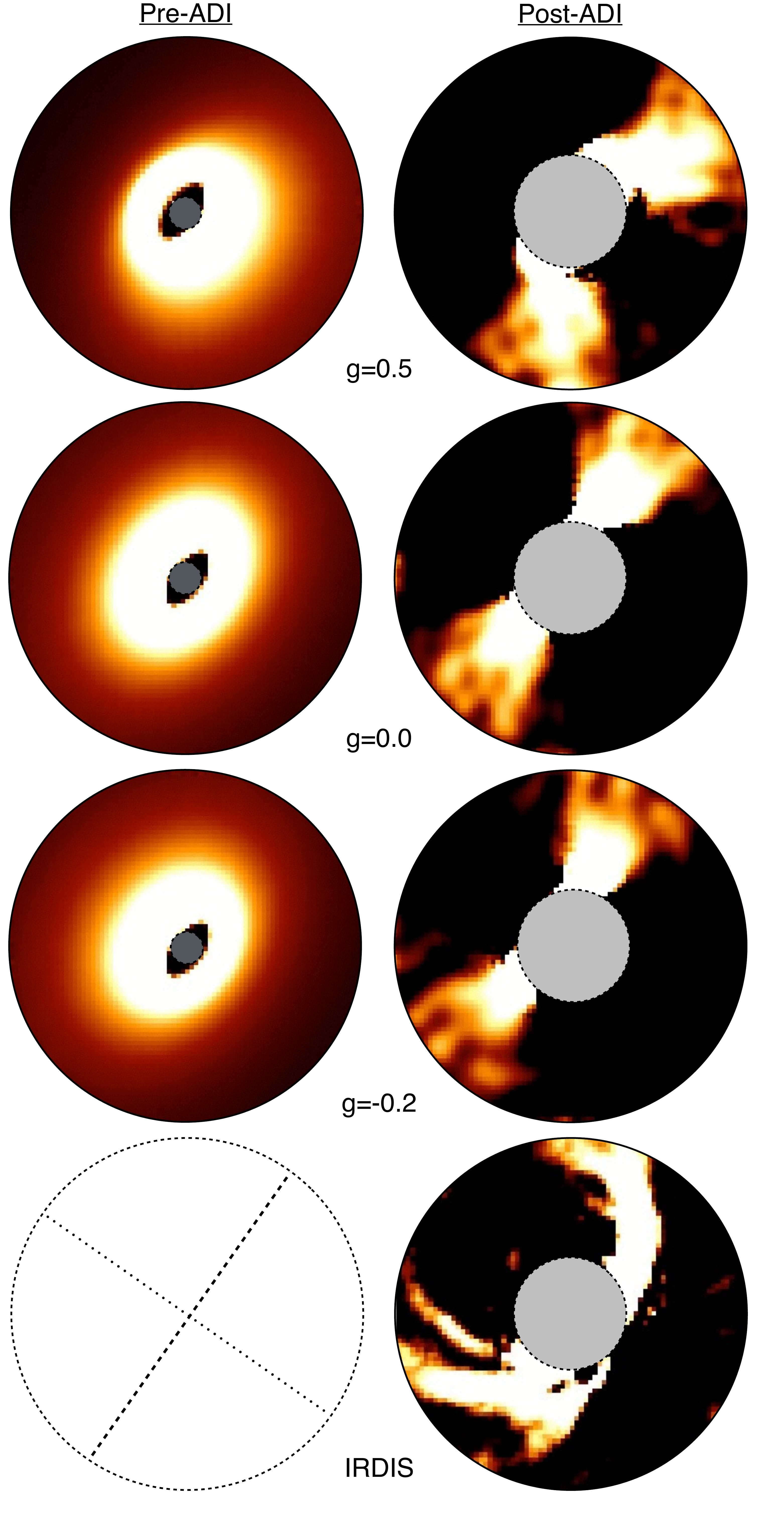}
  \caption{The impact of ADI processing on the scattered light distribution from an inclined disk. Left column: synthetic scattered light images in the H band obtained from the HD100546 model by \citet{Mulders2013a}. Right column: the same after the ADI analysis performed by \textsc{pynpoint}. The different rows are models with different asymmetry parameters $g$ with the last one being the IRDIS observations. The flux scales are arbitrary.}
            \label{ADI}
  \end{figure}

\textit{Structures in the ADI images.} The brightness distribution from the IRDIS images significantly differs from that of polarized light. From the small and medium scale images of Fig.\ \ref{Irdis_imagery}, it appears as a double-wing structure rising from the major axis and arching toward the far side \citep[similarly to the GPI images by][]{Currie2015}. Contrarily to the polarized light, the region between the two wings (to NE) is dark. To explore to what extent this morphology can be generated by the ADI processing, we produced synthetic images of HD100546 in the H band from the model by \citet{Mulders2013a} with Henyey-Greenstein asymmetry parameter $g$ \citep{Henyey1941} spanning {the range} from 0.5 to -0.2. Then, we convolved these images with the angular resolution of the IRDIS observations, replicated and rotated them to reproduce the field rotation of our observations (see Sect.\ \ref{IRDIS}) and gave them as input to \textsc{pynpoint}, which was {run} with the same setup described in Sect.\ \ref{IRDIS} (with the small scale boundaries). Some illustrative outputs of this exercise are shown in Fig.\ \ref{ADI}. It turned out that the ADI processing transforms a continuous brightness distribution into a double-wing structure, with the wings always rising from the major axis and the azimuthal distance between them being strongly dependent on $g$. None of the test models could reproduce the observed morphology. Speculatively, this supports the idea of a strong backward-peaking phase function. However, a deeper discussion is beyond the scope of the current paper. Broadly speaking, this double-wing structure is a new form of feature \citep[see also AK Sco,][]{Janson2016} that may be recurrent among the observations of inclined disks carried-out with the new generation AO systems. Our exploration has shown that such a peculiar structure may still be interpreted in the context of a disk with an azimuthally continuous brightness distribution.

\subsection{Disk structures} \label{Discussion_structures}
Almost all protoplanetary disks imaged so far with high resolution show peculiar features. Spirals \citep[e.g.,][]{Muto2012, Garufi2013, Benisty2015} and annular gaps \citep[e.g.][]{Debes2013, Quanz2013b, Rapson2015} are, among these structures, the most intriguing. However, none of these disks seem to show both types of structures. The disk around HD100546 only shows spiral arms. Nonetheless, the absence of annular gaps is not less important in the context of planet-disk interaction and is discussed in Sect.\ \ref{Discussion_b}. 

The inner spiral arm at $\sim 0.2\arcsec$ \citep[see Fig.\ \ref{Imagery}e and][]{Avenhaus2014b} is evident in polarized light but is not in the ADI images. This non-detection is most likely due to the ADI processing, which tends to cancel out any disk feature with high azimuthal symmetry and to over-subtract flux along the disk minor axis (see Fig.\ \ref{ADI}). The aperture of this spiral significantly differs from that of all others. In fact, if we define the aperture $a$ as from $r=a\theta$ (with $r$ distance from the star in arcseconds and $\theta$ azimuthal angle in radians\footnote{To obtain $a$, we de-projected the disk following \citet[$i=42\degree$ and ${\rm P.A.}=145\degree$]{Pineda2014} and considered a disk opening angle of 10$\degree$.}), this spiral is fitted by $a \sim 0.2$ whereas the other features farther out show $a$ as high as $0.5-0.6$. The most interesting finding about this spiral is the absence of rotation on a nine-years timescale (see comparison with NACO data in Fig.\ \ref{Comparison}c). This is surely not consistent with a Keplerian motion around the star. In fact, a spiral at a de-projected separation of 30 AU from a 2.4 $\rm M_\sun$ star \citep{vandenAncker1997}, has an orbital period of roughly 100 years, which yields an expected rotation of about $30\degree$ in nine years. 

The outer Southern spirals from IRDIS also show no apparent motion with respect to the detection by \citet{Boccaletti2013} (see Fig.\ \ref{Comparison}d). Given the 5.2 years between the two epochs and a de-projected radius of 120 AU, the Keplerian motion of these spirals projected onto the disk plane is $\sim 1.5\degree$, which translates into 0.04\arcsec \ at those radii. It is difficult to determine the exact precision of our comparison because of the ADI processing. However, a visual inspection of the relative position of the spirals yields no sign of such a shift, as the edges of these features (whose width is on average 0.06\arcsec) match very well between the two datasets.  

The explanation for the absence of rotation is not obvious. If we focus on the inner spiral only (where the deviation from Keplerian motion is safe), one could try to relate its motion to that of a perturber object at larger radii. In fact, if this feature is excited by a massive object orbiting farther out (see discussion below), the spiral will be locked to the Keplerian motion of such an object. By conservatively assuming that the spiral has moved by less than $10\degree$ in nine years, this scenario would point toward a perturber object at more than 70 AU. Alternatively, the observed feature may only resemble a spiral arm and instead be a geometrical property of the disk (such as the top of the inner wall) whose appearance in scattered light is, therefore, not supposed to change with time. Analogously, one may be tempted to explain the Southern spirals with the geometrical properties of the disk. If one gives credence to the Western spirals of Fig.\ \ref{Irdis_imagery}, the global arm-pattern visible from the near side resembles the wing-structure observed at small radii, with multiple arms symmetric around the minor axis which arch toward the far side. In this scenario, the Northern spirals would be morphologically different from the Southern/Western spirals and possibly be in Keplerian rotation. Such an idea can be tested {after} 2020.

The interpretation of the multiple arms from the IRDIS small/medium scale (left/middle panels in Fig.\ \ref{Irdis_imagery}) is also not straightforward. The perfect symmetry around the minor axis suggests that this pattern is connected to the dust scattering properties and to the ADI processing from an inclined disk. As discussed in Sect.\ \ref{Discussion_morphology}, the morphology of the wing structure can be explained by the ADI processing. It is also nonetheless clear from Fig.\ \ref{ADI} that sharp features as the multiple arms observed from IRDIS cannot be generated from the pipeline. Moreover, the spatial consistency between different \textsc{pynpoint} runs, different instruments and observing modes reinforce that these features are real, even though their morphology differs from the \textit{nominal} spiral arms frequently observed from face-on disks \citep[e.g.,][]{Garufi2013, Wagner2015}. Second epoch observations of these inner features (as early as in 2018) and additional high-contrast ADI images as well as complementary RDI \citep[Reference star Differential Imaging,][]{Mawet2012} images of inclined disks will help unravel the nature of these arms.   


A lively debate on the nature of the increasingly observed spirals in protoplanetary disks is ongoing. First of all, it has not been solved yet whether the detection of these features in scattered light reflects an intrinsic change in the dust distribution/properties down to the disk mid-plane \citep[as proposed for SAO206462 by][]{Perez2014, Quanz2015b}. Many authors instead favor a scenario where these are due to a change in the pressure scale height \citep[e.g.,][]{Juhasz2015, Pohl2015}. Disentangling the spiral morphology is also fundamental to determine their causes. The effects of the gravitational instability \citep{Durisen2007} on the disk morphology has been studied with hydrodynamical simulations by e.g., \citet{Dong2015b}, which successfully reproduce the appearance of SAO206462 and MWC758. However, gravitational instability has been disfavored in many specific cases because of the insufficiently massive nature of these disks \citep[e.g.,][]{Boccaletti2013, Garufi2013}, even though the gas mass in these disks is not firmly constrained. Recently, planet-disk interactions have been {investigated quantitatively} \citep[e.g.][]{Juhasz2015, Pohl2015, Dong2015a}. Despite the good agreement with the observed brightness contrast of spirals, {this hypothesis} has difficulties in reproducing the observed aperture of spirals (unless planets at very large radii or very high disk scale heights are invoked). Thus, the current framework does not allow us to firmly ascribe the spiral structure of HD100546 to any scenario.

\subsection{Disk interaction with planet b} \label{Discussion_b}
Independent multi-epoch and -filter observations of HD100546 in thermal IR have revealed the existence of a point source at $r \sim 0.46\arcsec$, sitting on top of an extended component \citep{Quanz2013a, Quanz2015, Currie2014, Currie2015}. The current interpretation of the point source is thermal emission from a young planet (possibly surrounded by a circumplanetary disk) which is released from an effective area with $R \sim 7 \ {\rm R_{jup}}$ and effective temperature $T_{\rm eff} \approx 930$ K \citep{Quanz2015}. This claim {implied} the rejection of a scattered light nature for this emission. The absence of a localized brightness enhancement from our ZIMPOL data in correspondence of the planet supports the validity of this rejection. Nonetheless, the large-scale emission around the point source is spatially consistent with our diffuse polarized emission to North (see Fig.\ \ref{Comparison}b), suggesting that {our observed} diffuse emission is (mostly) scattered light. A similar feature is detected by \citet{Quanz2015} and \citet{Currie2014} to the SE and it also matches the diffuse enhancement from ZIMPOL.

Our ADI K band image reveals an extended emission at the location of the \textit{b} planet (see mid-bottom panel of Fig.\ \ref{Irdis_imagery}). This emission has no equivalent in the H band, and such a discrepancy is the only significant difference between the two wavebands. This dissimilarity can be either explained by the different optical depth of incident light or by the different extinction experienced through the disk by a local emission. The former explanation implies that the emission is purely scattered light and that a change in the disk morphology occurs between the depths where photons with H and K wavelengths are scattered. However, these two surfaces are probably too close to allow such a change. Moreover, the lack of a similar discrepancy between the H and K$_{\rm S}$ band NACO images by \citet{Avenhaus2014b} casts further doubt on the hypothesis. The latter explanation lies on the assumption that a local source of heat is present at a certain disk depth. This thermal emission would be subject to a higher extinction at shorter wavelength and would thus explain the observed difference. The recovery of this emission might still be possible in the H band, as shown by \citet{Currie2015}. In any case, it remains to be understood whether the K band detection is actually related to any process of planet formation. A more in depth analysis of this emission, complemented by the simultaneous IFS data, will be presented by Sissa et al.\ (in prep.).

From the entire dataset analyzed in this paper, any disk interaction with planet \textit{b} remains fairly elusive. First of all, there is no hint of any disk gap at the planet location. Based on the local noise of our images, we can rule out with 3$\sigma$ confidence any brightness discontinuity more pronounced than $16\%$. We also integrated the flux from the back side contained in an ellipse passing through \textit{b} (obtained consistently with the disk geometry) but we could not infer any discontinuity. The absence of any brightness discontinuity does not necessarily imply the absence of gaps {in the distribution of $\mu$m-sized particles}, since the high inclination of the source may not enable the observer to see through a narrow hole. From the angular resolution of the ZIMPOL data ($\sim 2$ AU), some naive geometrical considerations, and the assumption of a 5 AU disk scale height at 50 AU \citep{Montesinos2015}, one obtains that our images could still detect a gap {in the small dust grain distribution} larger than $\sim 7$ AU. Furthermore, the observation of a significant portion of the disk wall at the outer gap edge may result in a significantly different flux distribution (as shown for the inner rim in Fig.\ \ref{Profile}a). The absence of an observed discontinuity is difficult to reconcile with the large {gaseous} gap expected to be sculpted by a giant planet in a few tens orbits timeframe \citep[e.g.,][]{Crida2006, Masset2008}, unless the planet is very young ($\sim 10^{4-5}$ years, given its orbital period). { \citet{deJuanOvelar2013} showed that small polarized light brightness discontinuities in the R band are expected from interactions with $> 1\, {\rm M_{jup}}$ mass planets}.  An additional possibility to keep in mind is that the polarized signal deriving from the disk surface is blended with the contribution from halo particles. Speculatively, this contribution may be significant in highly inclined systems as suggested by the notion that sharp, prominent features like spirals and annular gaps are hardly observed in polarized light from inclined disks \citep[see e.g.,][]{Hashimoto2012, Takami2013, Follette2015, Thalmann2015}.        

The only disk properties that can be ascribed to the interaction with \textit{b} are the polarized light peak at comparable azimuthal angles (see Fig.\ \ref{Profile}d) and the Northern arm highlighted in Fig.\ \ref{Imagery}e and Fig.\ \ref{Irdis_imagery}. As discussed in Sect.\ \ref{Discussion_morphology}, the large-scale enhancement of scattered light to the North is a possible imprint of a diffusely higher disk scale height. This may be in turn due to the hot environment produced by an enhanced disk accretion rate in correspondence of a luminous giant planet \citep[as shown by the hydrodynamical simulations customized to HD100546 by][]{Montesinos2015}. Finally, the spatial connection between the northern arm detected in both ZIMPOL and IRDIS and the planet \textit{b} is tantalizing. As discussed in Sect.\ \ref{Discussion_structures}, spiral arms are a natural outcome of the planet-disk interaction. However, this arm appears wrapped in the opposite direction of the other spirals and, as commented in Sect.\ \ref{Discussion_structures}, it is not clear yet to what extent this and other arm-like features in our datasets can be due to a combination of dust scattering properties and disk geometry.

\section{Summary and conclusions} \label{Conclusions}
We present the first SPHERE observations of HD100546 in scattered light, obtained with the scientific sub-systems \mbox{ZIMPOL} (polarized R$'$ band) and IRDIS (H and K band). Complementary data from previous works enabled us to draw a comprehensive picture of the disk emission from 0.6 $\mu$m to 4.8 $\mu$m.

The ZIMPOL images in the visible resemble previous images in the near-IR, with the presence of an ellipsoidal cavity, of two bright lobes at the disk inner rim and of a tremendous brightness contrast between the disk near (dark) and far side (bright). The main findings from the analysis of these images are:

\begin{itemize} 

\item The cavity size in the visible is $11 \pm 1$ AU. This is consistent with the estimate from mid-IR interferometry by \citet{Panic2014}. This finding implies that the CO emission associated to the planet candidate \textit{c} \citep{Brittain2013} is located \textit{within} the disk. The marginal difference with the cavity size in the near-IR (13 AU) can be ascribed to the different disk opacity at those wavelengths. The large discrepancy with the cavity size at millimeter wavelengths \citep[25 AU,][]{Wright2015} is qualitatively consistent with the dust differentiation expected from the interaction with a (yet unseen) giant planet in the disk cavity. The amount of such a discrepancy requires a $\gtrsim 15~M_{\rm Jup}$ mass companion.

\item A luminous wedge stands out from the dark side and this is much brighter than {seen in} previous near-IR images. The cause of this difference may also clarify the nature of the global darkness. If it is a time-related difference, then the wedge might be a penumbra in a globally shadowed region. If it is a wavelength-related difference, then we may be tracing the forward peak of an otherwise backward peaking scattering phase function. We favor the latter scenario and entrust the answer to upcoming near-IR images and custom models.

\item The spiral arm at $r \sim 30$ AU revealed by \citet{Avenhaus2014b} does not show any proper motion in a nine-years timescale, inconsistent with being in Keplerian motion. This could either mean that the spiral is locked with the orbital motion of a companion at $\gtrsim 70$ AU or that it is an azimuthally symmetric feature misinterpreted as a spiral.  

\item The effect of multiple scattering might be {important}. This is suggested by the morphology of the U$_\phi$ images, quantitatively resembling the expectations for an optically thick, significantly inclined disk when this phenomenon is taking into account \citep{Canovas2015}.

\end{itemize}

The near-IR IRDIS images, analyzed in ADI with \textsc{pynpoint}, show a complex multiple-arm structure at all spatial scales. In the H band, a bright knot along one of these arms roughly matches the location of the putative detection of planet \textit{c} by \citet{Currie2015}. This knot is not present in the K band. Moreover, an extended emission in the K band is detected at the location of the \textit{b} planet \citep{Quanz2013a}. Our conclusions on the analysis of these images are:

\begin{itemize}

\item The double-wing structure symmetric around the minor axis can be generated from a continuous disk by the ADI processing. However, the multiple-arm morphology is consistent through different reductions and instruments and is therefore real. It is yet unclear whether these features are all spiral arms at the disk surface or marginal anisotropies accentuated by the scattering properties from an inclined disk. Their temporal evolution will provide fruitful insight into their nature. 

\item The extended brightness associated to HD100546b is most likely a thermal emission originating deep in the disk. The non-detection in the H band (and from all polarized light datasets) is in fact difficult to {reconcile} with the scattered light scenario. An in-depth analysis of this detection will be described by Sissa et al.\ (in prep.).

\end{itemize}

The direct comparison between the ZIMPOL and the IRDIS images with previous works also provide some insight into the disk geometry and the scattering properties. The difference between the total and the polarized light in the H band can be mainly ascribed to the ADI processing. However, the different brightness distribution at the disk inner edge may indicate that particles at the inner rim are more prone to forward-scatter photons, contrarily to what is seen at larger radii. Furthermore, the overall similarity of the polarized light at 0.6 $\mu$m and the total intensity at 4.8 $\mu$m supports the idea that the latter emission is also (mainly) scattered light.  

All things considered, the imprints of the (probable) giant planets around HD100546 on the disk morphology remain elusive. We found no strong evidence of any disk feature which might be due to the interaction with HD100546c. Some of them (i.e.\ an arm apparently originating along the major axis, the enhanced polarization of a lobe at the inner rim, and a spiral arm at 30 AU) are, most likely, fortuitous. On the other hand, the maximized polarized brightness on a large scale around HD100546b may be speculatively connected to the expected enhanced disk accretion rate in proximity of the planet. Furthermore, the spatial connection between an arm-like structure and the localized thermal emission in the K band may also reveal some yet unclear planet-disk interplay. The absence of a detectable gap in correspondence of planet \textit{b} also raises unanswered questions on the processes of planet formation. A narrow ($\lesssim 7$ AU) or a shallow (providing a $\lesssim 16\%$ flux decrease) gap, as well as a very young nature for \textit{b} may reconcile with our observations. 

These data are a showcase of the capability of the new generation high-contrast imager SPHERE. However, a lot remains to be understood about the enigmatic {planet-forming} disk of HD100546. The future confirmation (or rejection) of planet candidates and the recovery of disk features similar to  those shown here from other suspected planet-forming disks (with SPHERE, ALMA, or GPI) will also help unravel the mysterious geometry of the disk around HD100546 and the mechanisms governing planet formation.
  
\begin{acknowledgements}
We thank Paola Pinilla and Sean Brittain for the fruitful scientific discussion, and the referee for a thorough review that helped improve the text. We are grateful to the ESO technical operators at the Paranal observatory for their valuable help during the observations. Part of this work has been carried out within the frame of the National Centre for Competence in Research PlanetS supported by the Swiss National Science Foundation. SPQ, HMS, and MRM acknowledge the financial support of the SNSF. HA acknowledges support from the Millennium Science Initiative (Chilean Ministry of Economy), through grant "Nucleus P10-022-F" and financial support from FONDECYT grant 3150643. This research has made use of the SIMBAD database, operated at the CDS, Strasbourg, France. SPHERE is an instrument designed and built by a consortium consisting of IPAG (Grenoble, France), MPIA (Heidelberg, Germany), LAM (Marseille, France), LESIA (Paris, France), Laboratoire Lagrange (Nice, France), INAF - Osservatorio di Padova (Italy), Observatoire de Geneve (Switzerland), ETH Zurich (Switzerland), NOVA (Netherlands), ONERA (France) and ASTRON (Netherlands) in collaboration with ESO. SPHERE was funded by ESO, with additional contributions from CNRS (France), MPIA (Germany), INAF (Italy), FINES (Switzerland) and NOVA (Netherlands). SPHERE also received funding from the European Commission Sixth and Seventh Framework Programmes as part of the Optical Infrared Coordination Network for Astronomy (OPTICON) under grant number RII3-Ct-2004-001566 for FP6 (2004-2008), grant number 226604 for FP7 (2009-2012) and grant number 312430 for FP7 (2013-2016).
\end{acknowledgements}

 \bibliographystyle{aa} 
\bibliography{Reference.bib}

\end{document}